\documentclass[11pt]{article}

\usepackage{amsmath,amsfonts}
\usepackage{bm}
\usepackage{natbib,theorem}
\usepackage[dvipdf]{graphicx}
\usepackage{amssymb}
\usepackage{color}
\usepackage{latexsym}
\usepackage{multicol}
\usepackage{epsfig}
\usepackage{setspace}
\usepackage{multirow}
\usepackage{hyperref}

\renewcommand{\baselinestretch} {1.30}
\newtheorem{proposition}{Proposition}

\theoremheaderfont{\hskip\parindent\normalfont}
\theorembodyfont{\normalfont}

\newcommand{\bP}{\mathbf{P}}

\newcommand{\bt}{\mathbf{t}}

\newcommand{\bX}{\mathbf{X}}

\newcommand{\bY}{\mathbf{Y}}

\newcommand{\cA}{\mathcal{A}}
\newcommand{\cB}{\mathcal{B}}

\newcommand{\cN}{\mathcal{N}}

\newcommand{\bbeta}{\boldsymbol{\beta}}

\newcommand{\btau}{\boldsymbol{\tau}}
\newcommand{\btheta}{\boldsymbol{\theta}}

\newcommand{\bzero}{\boldsymbol{0}}
\newcommand{\bone}{\boldsymbol{1}}

\def\v{{\varepsilon}}
\def\bv{{ \boldsymbol{\varepsilon}}}

\newcommand{\argmax}{\operatornamewithlimits{argmax}}
\newcommand{\argmin}{\operatornamewithlimits{argmin}}

\newcommand{\red}{\textcolor{black}}
\newcommand{\blue}{\textcolor{black}}

\begin{document}

\renewcommand{\baselinestretch}{1.1}

\title{Multiple Change-point Detection: a Selective Overview}

\author{Yue S. Niu, Ning Hao, and Heping Zhang\\
University of Arizona and Yale University}
\date{\today}
\maketitle

\begin{abstract}
Very long and noisy sequence data arise from biological sciences to social science including high throughput data in genomics and stock prices in econometrics. Often such data are collected in order to identify and understand shifts in trend, e.g., from a bull market to a bear market in finance or from a normal number of chromosome copies to an excessive number of chromosome copies in genetics. Thus, identifying multiple change points in a long, possibly very long, sequence is an important problem.  In this article, we review both classical and new multiple change-point detection strategies. Considering the long history and the extensive literature on the change-point detection, we provide an in-depth discussion on a normal mean change-point model from aspects of regression analysis, hypothesis testing, consistency and inference. In particular, we present a strategy to gather and aggregate local information for change-point detection that has become the cornerstone of several emerging methods because of its attractiveness in both computational and theoretical properties.
\end{abstract}

\section{Introduction}\label{S1}
Studies of change-point detection problem date back to 1950s \citep{page1954continuous,Page:1955,Page:1957}. Since then, this topic has been of interest to statisticians and researchers in many other fields such as engineering, economics, climatology, biosciences, genomics and linguistics, just to name a few. In many applications, observed are an ordered sequence of random quantities, from which the change points, i.e., positions of structural change, are inferred. Examples of such sequences include the daily average temperatures of a specific location over the years, the quantity of some harmful elements, e.g., antimony, in the drinking water, stock prices at some time points over a period, and recently, sequencing data in genomics. Depending on the goal of the data that are collected, detection of change points can be crucial for decision making or necessary for understanding certain scientific issues. \blue{Many methods} have been introduced to detect the change of the mean, variance, slope of regression line, hazard rate, or nonparametric distribution for various models. We refer to the books by \cite{brodsky1993nonparametric,brodsky2000non,carlstein1994change,CsorgoHorvath:1997,ChenGupta:2000:book} for various aspects for classical change-point analysis, and to a recent article \cite{lee2010change} for a list of comprehensive bibliography of books and research papers on this topic. While this body of work constitutes a rich literature, it mainly deals with the inference of a single change in a short or moderate sized sequence. Detecting multiple change points in a very long sequence has emerged as an important problem that has attracted more and more attention recently.

We review both classical and new multiple change-point detection strategies and discuss their strengths and limitations by examining the general strategies, assessing the computational complexity, and establishing the asymptotic theory. In particular, we present the distinctive characteristic of multiple change-point model from the single change-point model, and give insights on the strategies employed by state-of-the-art change-point detection methods. Some interesting research directions are also discussed. We should point out that there exist a massive number of research papers on change-point detection or closely related topics. We \blue{will concentrate mainly on frequentist approach and} narrow our review and discussion to the so called ``posteriori change-point detection'' problem according to the terminology of \cite{fryzlewicz2012wild}. Even with this focus, it is impossible to do justice for all related work, and hence we can offer only a selective overview.

The rest of paper is organized as follows. Section \ref{S2} introduces several formulations of multiple change-point model and its distinct features. Section \ref{S3} reviews a variety of multiple change-point detection techniques. Section \ref{S5} focuses on theoretical properties of these methods. Section \ref{S6} summarizes new strategies used by some recent methods. We end this paper with some concluding remarks in Section \ref{S7}.


\section{Multiple change-point problem}\label{S2}

\subsection{From classical to modern data}
\begin{figure}  
\begin{center}
\includegraphics[width=5.2in,height=3.8in]{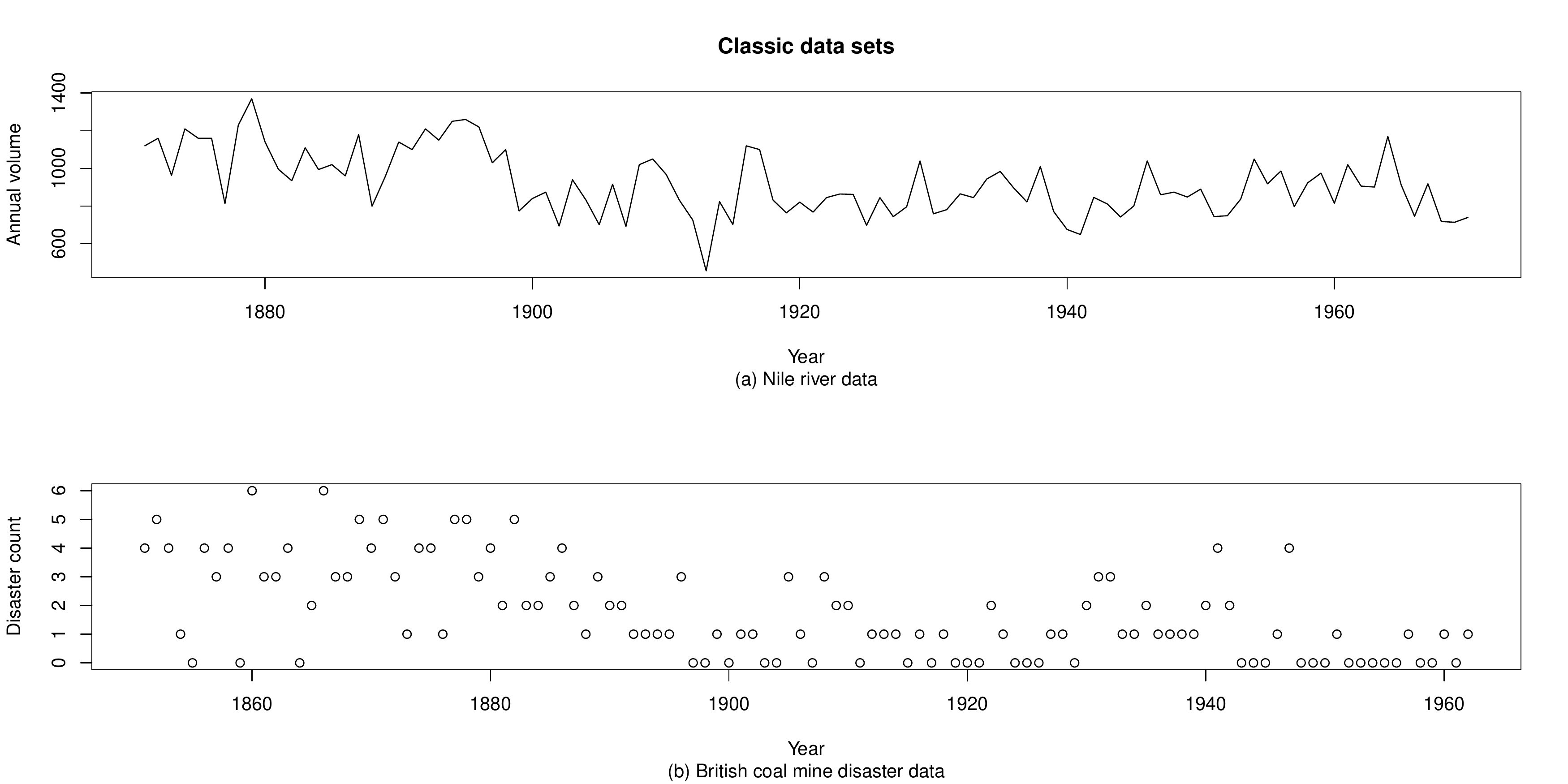}                    
\end{center}
\caption{Two classical data sets. (a) Measurements of the annual flow of the river Nile at Ashwan 1871-1970. (b) Yearly counts of British coal-mining disasters between 1851-1962.}
\label{fig1}
\end{figure}

\begin{figure}  
\begin{center}
\includegraphics[width=5.2in,height=3.2in]{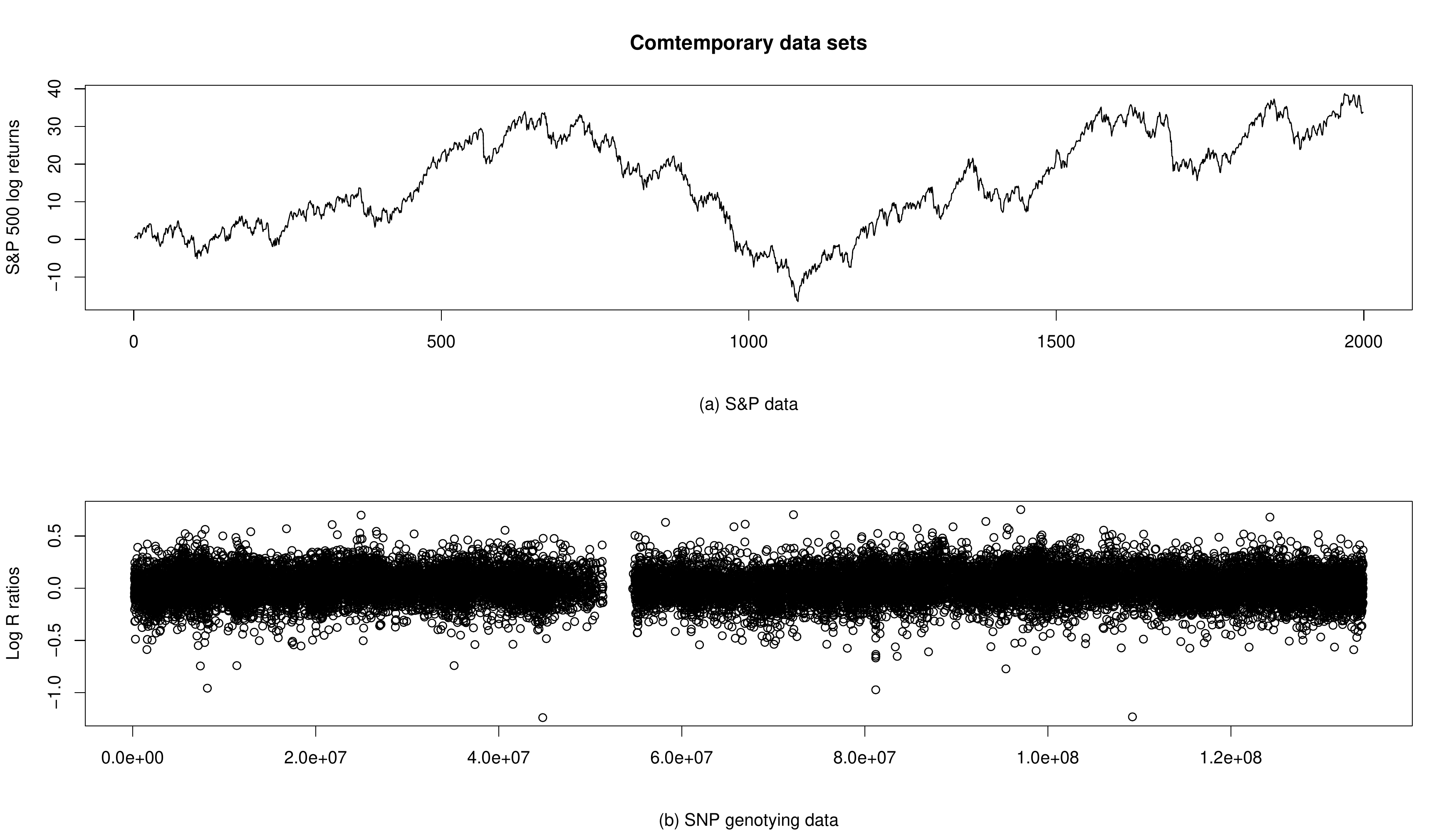}                    
\end{center}
\caption{Two contemporary data sets. (a) Log-returns on the daily closing values of S\&P 500 of length 2000, ending 26 October 2012. (b) Log R ratios of length 27272 along Chromosome 11 of the subject father from a SNP genotying data set for a father-mother-offspring trio produced by Illumina 550K platform, downloaded from \texttt{http://www.openbioinformatics.org/penncnv/}.}
\label{fig2}
\end{figure}

Two datasets, Nile river data \citep{cobb1978problem} and British coal mine disaster data \citep{carlin1992hierarchical}, plotted in Figure 1 (a) and (b), respectively, have been commonly used in classical change-point analyses. The main focus of those analyses was to test whether there was a change point along the sequence. Work on detection for multiple change points began in 1980's (e.g.,  \cite{Vostrikova:1981,Yin88,Yao1988,Yao:1989}), although the applications seemed limited during that decade. The advent of high-throughput technologies has produced high dimensional data that are of great interest in statistical sciences, and not surprisingly, revitalized the change-point analysis. In Figure \ref{fig2}, we plot a time series data set and a SNP array data set, which represent typical modern applications of multiple change-point model. A key difference between Figures \ref{fig1} and \ref{fig2} is the number of data points.

\subsection{A variety of multiple change-point model formulations}
Let $\bY=(Y_1,...,Y_n)^{\top}$ be a sequence of independent random variables (or vectors) with probability distribution functions $F_1$, ..., $F_n$, respectively. A change point occurs at $\tau$  when $F_{\tau}\ne F_{\tau+1}$. The goal is to estimate the number of change points, $J$, and location vector of change points $\btau=(\tau_1,\cdots,\tau_J)^{\top}$. We use the convention  $\tau_0=0$ and $\tau_{J+1}=n$ throughout this paper. Depending on applications, we may assume that the distributions $\{F_i\}_{i=1}^n$ belong to a common parametric family or keep the setting nonparametric.

The simplest case where $J\le 1$ has been studied extensively. It is usually formulated as a hypothesis testing problem,
\begin{eqnarray}\label{aaa1}
H_0 &:& F_1=F_2=\cdots=F_n \text{  versus}\\
\nonumber H_1 &:& F_{1}=\cdots=F_{\tau}\ne F_{\tau+1}=\cdots=F_{n} \text{ with unknown } \tau.
\end{eqnarray}
The first step is to test hypotheses $J=0$ versus $J=1$, and if we reject $J=0$, the next step is to make inference on the location parameter $\tau$ \citep{Hawkins:1977}.

Both steps turned out to be \blue{much more} complicated than they appeared. For example, even if $\{F_i\}_{i=1}^n$ are Gaussian with equal variance, the exact distribution of the likelihood ratio test statistic under $H_0$ is formidable, and usually approximated by asymptotic or numerical methods \citep{SenSri:1975,Hawkins:1977,CsorgoHorvath:1997}. The inference on $\tau$ has been regarded as a difficult problem, and studied in \cite{hinkley1970inference,worsley1986confidence,siegmund1988confidence} for normal mean model and one-parameter exponential families.

Concerning $J>1$, a normal mean change-point model is typically employed, and here we focus on this model. First of all, it is the most basic multiple change-point model (MCM). In fact, even under this basic model, there are many challenging and open problems. Second, despite the simplicity, it is useful in many applications including genomics and econometrics; see \cite{BBM:2000,OVLW:2004,ZhangSiegmund:2007,NiuZhang:2010,frick2014multiscale,fryzlewicz2012wild}, among many others.  Third, detecting changes in a sequence can often be reduced to the detection of mean changes in another derived sequence \citep{carlstein1994change,brodsky2000non}. Last but not least, it serves as a benchmark model for comparing different methods of detecting change points \citep{fryzlewicz2012wild}.
Next we discuss a few formulations of the normal mean MCM.

\subsubsection{Normal mean multiple change-point model}
Consider a sequence of independent random variables $Y_1,\cdots,Y_n$ with $Y_i\sim\cN(\theta_i,\sigma^2)$ and the mean parameter shifts at $J$ unknown locations as introduced above.

The normal mean MCM can be presented as
\begin{eqnarray}\label{aaa2}
Y_i=\theta_i+\v_i, \qquad \v_i\stackrel{iid}{\sim}\cN(0,\sigma^2), \qquad 1\leq i\leq n, \quad \text{ with}
\end{eqnarray}
\[ \theta_1=\theta_2=\cdots=\theta_{\tau_1}\neq\theta_{\tau_1+1}=\cdots=\theta_{\tau_2}\neq\theta_{\tau_2+1}=\cdots\quad\cdots =\theta_{\tau_J}\neq\theta_{\tau_J+1}=\cdots=\theta_n.\]

In other words, we assume that $\btheta=(\theta_1,...,\theta_n)^{\top}$ is piecewise constant with jumps or drops at $\btau=(\tau_1,...,\tau_J)^{\top}$.

Besides the piecewise structure, the change points are usually assumed to be sparsely located along the sequence. Roughly speaking, it means that $J$ is much smaller than $n$ and change points are not too close to each other. Therefore, we may view (\ref{aaa2}) as a high dimensional sparse model. Moreover, its natural sequential structure is a distinct feature and crucial in developing efficient algorithms.

\subsubsection{A regression model}\label{regmod}
Let
\[\beta_0=\theta_1, \quad \beta_j=\theta_{j+1}-\theta_{j};\quad j=1,...,n-1.\]
Then, model (\ref{aaa2}) is transformed to a linear model,
\begin{eqnarray}\label{aaa3}
Y_i=\sum_{j=0}^{i-1}\beta_j+\v_i,\qquad i=1,...,n,
\end{eqnarray}
or in matrix form
\begin{eqnarray}\label{aaa4}
\bY=\bX\bbeta+\bv,
\end{eqnarray}
where $\bbeta=(\beta_0,...,\beta_{n-1})^{\top}$ is coefficient vector with $J$ nonzero entries besides the intercept $\beta_0$ and the design matrix $\bX=(\bX_0,...,\bX_{n-1})$, with $\bX_j=(\bzero_{j}^{\top},\bone_{n-j}^{\top})^{\top}$. Here  $\bone_k$ and $\bzero_k$ are $k$-dimensional column vectors with equal entries $1$ and $0$, respectively. We see that $\bX$ is simply a lower triangular matrix with all nonzero elements equal to 1. Model (\ref{aaa3}) is sparse when $J$ is much smaller than $n$.

We can observe from
\begin{eqnarray*}
\btheta=\bX\bbeta
\end{eqnarray*}
that the columns  of $\bX$ spans the space of all piecewise constant \blue{vectors of length $n$}. Note that each column, say, $(\bzero_{j}^{\top},\bone_{n-j}^{\top})^{\top}$, has only one discontinuity at $j$. Therefore, the piecewise constant $\btheta$ is a linear combination of \blue{piecewise constant vectors with a single discontinuity}.

With this formulation, modern high dimensional sparse regression techniques can be applied to solve change-point problem directly; see e.g., \cite{HuangEtal:2005,fusedlasso:2008,Rinaldo:2009,ZhangLange:2010,harchaoui2010multiple,qian2012pattern}.

\subsubsection{Multiple hypothesis testing}\label{S2.2.3}
We can generalize the hypothesis testing problem for a single change in (\ref{aaa1}) to the multiple change-point setting as
\begin{eqnarray}\label{aaa5}
\quad H_0: \text{there is no change point} \quad \text{versus }\quad H_1:\text{ at least one change point}.
\end{eqnarray}
Unfortunately, the alternative hypothesis above is too broad to be useful in practice. To be specific about each location as a candidate of change, it becomes a multiple testing problem which tests whether or not each of individual positions is a change point. In other words, we test
\begin{eqnarray}\label{aaa6}
\quad H_0(j): j \text{ is a not change point; } \quad\text{versus} \quad H_1(j): j \text{ is a change point, }
\end{eqnarray}
where $j=1,2,...,n-1$.
In particular, for the normal mean model, $H_0(j)$ and $H_1(j)$ correspond to $\beta_j=0$ and $\beta_j\ne 0$ in model (\ref{aaa3}), respectively. This clearly involves a large number of tests. The existing methods and theory on the control of false discover rate \citep[FDR,][]{BH:95} can be applied if we ignore the the sequential structure in the data. However, if we want to take advantage of the sequential structure, \red{this is not a typical multiple testing problem}. \cite{HaoNiuZhang2013} attempted to address this problem by generalizing the concept of FDR and testing a series of window-shifting hypotheses:
\begin{eqnarray}\label{aaa7}
H_0(j)&:& F_{j+1-h}=\cdots=F_{j+h} \text{  versus}\\
\nonumber H_1(j)&:& F_{j+1-h}=\cdots=F_{j}\ne F_{j+1}=\cdots=F_{j+h},
\end{eqnarray}
where $h$ is a fixed integer with the assumption that any adjacent change points must be least $h$ points apart. \blue{In practice,} \red{we} \red{may determine $h$ when prior knowledge on the distances among the change points is available.} For the normal mean change-point model, (\ref{aaa7}) is reduced to
\begin{eqnarray}\label{aaa8}
&H_0(j):& \theta_{j+1-h}=\theta_{j+2-h}=\cdots=\theta_{j+h}; \quad\text{versus} \nonumber \\
&H_1(j):& \theta_{j+1-h}=\cdots=\theta_{j}\ne\theta_{j+1}=\cdots=\theta_{j+h}.
\end{eqnarray}
Note that the alternative hypothesis in (\ref{aaa7}) is more specific than that in (\ref{aaa6}).

\subsection{Locality and symmetry}\label{S2.3}

Relative to the single change point problem, MCM involves the unknown $J$ which creates \red{additional} difficulties \red{to} the case when $J$ is \red{known}. MCM is also distinct from the single change point problem in two other aspects.

The first one is the \emph{local nature} of the MCM. Consider a simple normal mean model with three change points at $\tau_1<\tau_2<\tau_3$ and assume that the error distribution is known. In order to infer $\tau_2$, the best way is to use only the data between $\tau_1$ and $\tau_3$ as other data are irrelevant. In practice, the error distribution is unknown, and the data outside the interval $(\tau_1,\tau_3]$ may be helpful only to learn the error distribution and find \red{a} threshold or stopping rule for some detection procedure. On the other hand, misuse of the information outside of the interval $(\tau_1,\tau_3]$ may bias the inference on $\tau_2$. Therefore, ideally, an oracle procedure would divide multiple change-point problem (\ref{aaa2}) into $J$ simple and local subproblems, which are single change-point problems over the segments $(0,\tau_2]$, $(\tau_1,\tau_3]$,..., $(\tau_{J-1},n]$, respectively. Although such a division is not feasible in practice, it is still helpful to mimic the oracle by using local information.

The second one is \emph{symmetry}. As in \cite{OVLW:2004}, we can connect the head and tail of the sequence to make it a circle. Now, the MCM becomes symmetric. As a result, there is no boundary effect and all locations play the same role. We should be cautious with this technique when $J$ is small, such as the single change point problem. The reason is that we might have \red{a} priori knowledge on whether there is a change point where the head and tail \red{connect}. \red{We} may lose such information using this circular model.
Nevertheless, in many applications it is harmless to consider this symmetric model. With this symmetry, it is interesting to study equivariant detection tools.

Both the difficulties and uniqueness of MCM make it a more interesting and challenging problem to investigate than the single change-point problem.

\subsection{Signal strength levels of MCM}\label{S2.4}

Change itself is a relative concept. Therefore, it is necessary for us to impose certain reasonable assumptions to avoid non-identifiability. Intuitively, there are two situations where it is difficult, if not impossible to detect a change point. The first case is when the mean shifts so small that we cannot distinguish between a real change or an effect of noise. The second scenario is when there are two (or more) change points \red{so} close to each other that we do not have enough data to draw inference. Therefore, it is reasonable to measure the overall signal strength of normal mean MCM by $S=\delta^2L$, where $L=\min_{0\leq \kappa\leq J}\{\tau_{\kappa+1}-\tau_{\kappa}\}$ is the minimal distance between change points and $\delta=\min_{1\leq i<n}\{|\theta_i-\theta_{i+1}|\}/\sigma$ is the ratio of the minimal shift to the noise level. 
\blue{We present in Table \ref{Tb:signal} some commonly used technical conditions. As we will see, the existing methods work best when these assumptions are satisfied.}

\begin{table}[hb]
\begin{center}
\caption{Different signal strength levels of MCM (\ref{aaa2}) used in the literature. Here $a_n\asymp b_n$ means $0<\frac{1}{M_1}<\frac{a_n}{b_n}<M_2<\infty$ as $n\to\infty$ for some constants $M_1$, $M_2$. $\|\cdot\|_{\infty}$ is infinity norm.}\label{Tb:signal}
\begin{tabular}{|l|l|l|}
  \hline
   Signal strength level& typical assumption & typical conclusion\\
   \hline
  S1. very strong on $L$      & (A1) $L\asymp n$, $\delta\asymp 1$ & (C1) $\|\hat\btau-\btau\|_{\infty}=O_P(1)$ \\
  S2. very strong on $\delta$ & (A2) $\delta^2\gg\log n$           & (C2) $\bP(\hat\btau=\btau)\to 1$ \\
  S3. strong                  & (A3) $\delta^2L \asymp \log n$     & (C3) $\delta^2\|\hat\btau-\btau\|_{\infty}=O_P(1)$ \\
  S4. weak                    & (A4) N/A                           & (C4) FDR control \\
  \hline
\end{tabular}
\end{center}
\end{table}
Early work such as \cite{Yao1988,Yao:1989} usually assumed that $\btau/n\to \bt$, a constant vector, and the mean shifts at change points and the noise variance $\sigma^2$ remain constants as $n\to\infty$. Under these assumptions, $\|\hat\btau-\btau\|_{\infty}=O_P(1)$, which is the optimal result (unless $\delta\to\infty$). An equivalent formulation is to assume that there is a fixed piecewise constant function on the interval $[0,1]$ to be estimated as the number of observations $n\to \infty$. This result extended the theory from the single change-point case to the multiple change-point case, but may not be applicable to some modern applications when the sequence is long and some change points are close to each other. \cite{Rinaldo:2009,qian2012pattern} used regression approach and obtained the consistency result $\bP(\hat\btau=\btau)\to 1$. But they required $\delta^2\gg\log n$, which may be unrealistic in practice. On the other hand, \cite{Donoho:Huo:2005} showed that $\delta^2L \geq 2\log n$ is a necessary condition to recover all change points even for an MCM with two change points. \cite{HaoNiuZhang2013} further showed that $\delta^2\|\hat\btau-\btau\|_{\infty}=O_P(1)$ under a slightly relaxed condition $\delta^2L \geq 32\log n$. 
\blue{It remains an open problem to find weakest possible sufficient and necessary conditions on $\delta^2L$ for full recovery of all change points.} When the signal is not strong enough for us to recover all change points, it is desirable to control the false positives.
As discussed by \cite{HaoNiuZhang2013}, it is challenging to establish a reasonable framework for this atypical multiple testing problem.


\section{Multiple change-point detection tools}\label{S3}
For multiple change-point  problem, the main goal is to estimate the number of change points and their locations. In this section, we review some classical and new approaches to identifying change points. From now on, we use $J^*$ and $\btau^*=(\tau^*_1,...,\tau^*_J)^{\top}$ to denote the true number of change points and their location (in ascending order) vector of a specific data generating process.

\subsection{Exhaustive search}
Ignoring its computational complexity, an exhaustive search among all possibilities $0\leq J\leq n-1$ and $0<\tau_1<\cdots<\tau_J<n$ can be applied. Define $\Gamma_J=\{\btau=(\tau_1,...,\tau_J):0<\tau_1<\cdots<\tau_J<n\}$ the set of all possible ordered $J$-dimensional vectors representing locations of $J$ change points.
For any $J$, define
\[\hat{\sigma}_J^2= \min_{\btau\in\Gamma_J}\hat{\sigma}_{\btau}^2,\]
where $\hat{\sigma}_{\btau}^2$ is the maximum likelihood estimator for variance conditional on the change-point location $\btau$. \cite{Yao1988} showed the consistency property of the estimator $\hat{J}$ determined by Bayesian Information Criterion (BIC)
\begin{eqnarray}\label{aaa9}
\hat{J}=\argmin_J \frac{n}{2}\log\hat{\sigma}^2_J+J\log n.
\end{eqnarray}
Furthermore, \cite{Yao:1989} showed the convergence rate of the location estimator
\begin{eqnarray}\label{aaa10}
\hat\tau_{\kappa}-\tau_{\kappa}^*=O_P(1),\quad 1\leq \kappa\leq J^*,
\end{eqnarray}
where $\hat\btau=(\hat\tau_1,...,\hat\tau_{J^*})^{\top}=\argmin_{\btau\in\Gamma_{J^*}} \hat{\sigma}_{\btau}^2$, and $J^*$ is the true number of change points, which can be estimated consistently by (\ref{aaa9}).

Exhaustive search among all possible subsets is not efficient from \red{a} computational point of view, and too intensive when $n$ is large. Making use of the sequential structure, dynamic programming techniques can be applied to reduce the computational burden down to the order of $O(n^2)$ \citep{BBM:2000,jackson2005algorithm}. \cite{killick2012optimal} further reduced the complexity to $O(n)$, under, however, an assumption which may not be practical in applications \citep{fryzlewicz2012wild}.

\subsection{Stepwise selection}
Stepwise selection is a popular substitute for the exhaustive search with reduced complexity and simple implementation. Forward and backward selection methods are well-known stepwise procedures in model selection. Both of them can be applied to the change-point problem. In particular, the forward selection method, which is called binary segmentation (BS), has been studied for a long time \citep{scott1974cluster,Vostrikova:1981}. Recently, circular binary segmentation \citep[CBS,][]{OVLW:2004} and wild binary segmentation \citep[WBS,][]{fryzlewicz2012wild} have been proposed enhance the power of BS in identifying short segments. In contrast, the backward selection procedure has not been widely used for change point detection with the exception of \cite{Shin2014}.

\subsubsection{Forward selection approach}

We now introduce the binary Segmentation algorithm \citep{scott1974cluster,Vostrikova:1981}. It is a forward stepwise method \red{with} the following steps.

Step 1: Test for no change point versus one change point (\ref{aaa1}). If $H_0$ is not rejected, stop. Otherwise, estimate the first change point $\hat{\tau}$, which divides the whole sequence into two segments;

Step 2: Test these two segments, respectively, for further segmentation;

Step 3: Repeat the procedure for each segment until no one can be segmented further.

We illustrate here a special BS algorithm for solving model (\ref{aaa2}) with known variance $\sigma^2=1$ based on the likelihood ratio method. BS tests the hypotheses recursively in each segment. The following is for the initial segment and the rest is similar.
\begin{eqnarray}\label{aaa11}
\nonumber H_0 &:&\theta_1=\cdots=\theta_n, \quad \text{versus}\\
H_1 &:&\theta_1=\cdots=\theta_j\neq\theta_{j+1}=\cdots=\theta_n \text{ for some } 1\leq j<n.
\end{eqnarray}
A likelihood ratio type statistics
\begin{eqnarray}\label{aaa12}
T_1=\max_{1\leq j\leq n-1}(-2\log\Lambda_j)
\end{eqnarray}
can be used \citep{SenSri:1975}, where
\begin{eqnarray}\label{aaa13}
-2\log\Lambda_j=(\bar{Y}_{j+}-\bar{Y}_{j-})^2/[j^{-1}+(n-j)^{-1}],
\end{eqnarray}
and $\bar{Y}_{j-}=\sum_{k=1}^j Y_k/j$ and $\bar{Y}_{j+}=\sum_{k=j+1}^n Y_k/(n-j)$.
Moreover, when the alternative is supported, the segment can be divided into two parts by $\hat j$, obtained from
\begin{eqnarray}\label{aaa14}
\hat{j}=\argmax_{1\leq j\leq n-1}(-2\log\Lambda_j).
 \end{eqnarray}
The distribution of $T_1$ under $H_0$ can be approximated numerically or asymptotically. We refer to \cite{CsorgoHorvath:1997} for asymptotic properties of $T_1$ and $\hat{j}$.

The main drawback of the BS algorithm is that it can \red{rarely} detect short segments embedded in the middle of long segments \citep{OVLW:2004,fryzlewicz2012wild,Shin2014}. 
To increase the power for the BS algorithm in detecting short segments, \cite{OVLW:2004} proposed the Circular Binary Segmentation (CBS) algorithm. \blue{The main idea is to splice two ends of the segment to make it a circle and check whether there exists a segment, say between $l$ and $r$, such that its mean is significantly different from the mean of the \red{remaining} part. In particular, they considered and applied recursively a test of no change against \red{a so-called} epidemic alternative, as described below.}
\begin{eqnarray}\label{aaa15}
\quad   H_0 &:&\theta_1=\cdots=\theta_n, \quad \text{versus}\\
\nonumber  H_1 &:&\theta_1=\cdots=\theta_l=\theta_{r+1}=\cdots=\theta_n\neq\theta_{l+1}=\cdots=\theta_r \text{ for a pair }l<r.
\end{eqnarray}
\red{Here the alternative hypothesis is called epidemic in analogy to an epidemic running from $l$ through $r$ after which the normal state is restored \citep{levin1985cusum,YaoQ:1993}.}
Similarly to (\ref{aaa13}), we can calculate the log likelihood ratio for a fixed pair $(l,r)$,
\begin{eqnarray}\label{loglikelihood2}
-2\log\Lambda_{l,r}=(\bar{Y}_{I}-\bar{Y}_{O})^2/[(r-l)^{-1}+(n-r+l)^{-1}],
\end{eqnarray}
where the outside mean $\bar{Y}_{O}=\sum_{k\leq l\text{ or } k>r} Y_k/(n-r+l)$ and inside mean $\bar{Y}_{I}=\sum_{k=l+1}^r Y_k/(r-l)$.

Therefore, one may use
\[T_2=\max_{1\leq l<r\leq n}(-2\log\Lambda_{l,r})\]
as a test statistic for problem (\ref{aaa15}).

\blue{Since the CBS algorithm considers all pairs of points when calculating the test statistic, it is more powerful in detecting the short segments and has become one of the most popular algorithms in some applications such as copy number variation detection. However, it is computationally more expensive than BS.}

Recently, \cite{fryzlewicz2012wild} proposed the wild binary segmentation (WBS) to improve BS algorithm. Instead of the global test (\ref{aaa11}), WBS considers the same test on a random segment $[s,e]$, i.e.,
\begin{eqnarray}\label{testSE}
\nonumber H_0 &:&\theta_s=\cdots=\theta_e, \quad \text{versus}\\
H_1 &:&\theta_s=\cdots=\theta_j\neq\theta_{j+1}=\cdots=\theta_e \text{ for some } s\leq j<e.
\end{eqnarray}
The test statistic $T^{[s,e]}_1$ and $\hat{j}^{[s,e]}$ are defined similar to $T_1$ and $\hat{j}$ in (\ref{aaa12}) and (\ref{aaa14}). WBS first draws random intervals $[s_m,e_m]\subset [1,n]$, $m=1,...,M$ and \red{defines} the first change point as
\[\hat{j}_{\text{\tiny WBS}}=\hat{j}^{[s_{\hat{m}},e_{\hat{m}}]},\text{ where } \hat m=\argmax_{1\leq m\leq M} T^{[s_m,e_m]}_1. \]
Then WBS repeats the same procedure on each segment until some stopping rule is met. WBS offers one way to localize the BS procedure using local and random intervals instead of the whole segment in segmentation.

\subsubsection{Backward selection approach}
\cite{Shin2014} considered a backward detection (BWD) procedure. It starts with $n$ groups, each of which contains only one data point, and then merges groups based on some criterion until a stopping time. This is similar to the bottom-up clustering analysis where small clusters in the lower levels are joined to form larger ones in the upper levels. Specifically,
\begin{enumerate}
\item Start with a superset $\mathbb{G}^{(0)}=\{\mathcal{G}^{(0)}_1, \mathcal{G}^{(0)}_2, \cdots, \mathcal{G}^{(0)}_n\}$ where $\mathcal{G}^{(0)}_i=\{i\}$.
\item At step $k$ given $\mathbb{G}^{(k-1)}=\{\mathcal{G}^{(k-1)}_1, \mathcal{G}^{(k-1)}_2, \cdots, \mathcal{G}^{(k-1)}_{n-k+1}\}$, a decision is made to merge two consecutive groups in $\mathbb{G}^{(k-1)}$. Say, if $\mathcal{G}^{(k-1)}_i$ and $\mathcal{G}^{(k-1)}_{i+1}$ are merged to one group, then in $\mathbb{G}^{(k)}$, \[\mathcal{G}^{(k)}_j=\left\{
                                                                                                                                         \begin{array}{ll}
                                                                                                                                           \mathcal{G}^{(k-1)}_j, & \hbox{if } j<i; \\
                                                                                                                                           \mathcal{G}^{(k-1)}_j\cup\mathcal{G}^{(k-1)}_{j+1}, & \hbox{if } j=i; \\
                                                                                                                                           \mathcal{G}^{(k-1)}_{j+1}, & \hbox{if } j>i.
                                                                                                                                         \end{array}
                                                                                                                                       \right.\]

\item \red{Either} stop if a stopping criterion is met; \red{or} complete the whole process and then determine a best model along the sequence $\mathbb{G}^{(k)}$, $0\leq k\leq n$.
\end{enumerate}

For the normal mean model (\ref{aaa2}), \cite{Shin2014} employed a dissimilarity index to decide which pair of segments to merge in each step $k$. The dissimilarity index between two groups is defined as
\begin{equation}\label{aaa18}
 \mbox{DI}_j^{(k-1)}= \mbox{DI}(\mathcal{G}_j^{(k-1)}, \mathcal{G}_{j+1}^{(k-1)})=\frac{\left| \bar Y_j^{(k-1)}-\bar Y_{j+1}^{(k-1)} \right|}{ \sqrt{ |\mathcal{G}_j^{(k-1)}|^{-1}+ |\mathcal{G}_{j+1}^{(k-1)}|^{-1} }},
\end{equation}
where $\bar Y_j^{(k-1)}=\frac{1}{|\mathcal{G}_j^{(k-1)}|}\sum_{m\in\mathcal{G}_j^{(k-1)}}Y_m$, and $|\mathcal{G}_j^{(k-1)}|$ is the cardinality (i.e., size) of the set $\mathcal{G}_j^{(k-1)}$. Thus for $i=\argmin_j \mbox{DI}_j^{(k-1)}$, $\mathcal{G}^{(k-1)}_i$ and $\mathcal{G}^{(k-1)}_{i+1}$ are merged at step $k$ . In other words, at each step, BWD merges two consecutive groups with the minimal dissimilarity index.
In fact, (\ref{aaa18}) is essentially the log likelihood ratio test statistic for
\begin{eqnarray}\label{aaa19}
\nonumber H_0&:& \mathcal{G}^{(k-1)}_j\text{ and }\mathcal{G}^{(k-1)}_{j+1}\text{ have the same mean,} \quad \text{versus }\\
H_1&:&\mathcal{G}^{(k-1)}_j\text{ and }\mathcal{G}^{(k-1)}_{j+1}\text{ have different mean}.
\end{eqnarray}

\subsubsection{Summary}
The pros and cons of forward and backward stepwise procedures are analogous to those in model selection. In practice, there is little harm \red{in using both and then comparing results}.
The BS algorithm takes $O(n)$ operations for each step, and $O(\log n)$ steps in the worst scenario. So the total complexity is $O(n\log n)$. The CBS optimizes over all pairs of points so the complexity is increased by an order of magnitude. In general, BS and CBS are easy to implement, and early stopping rule can be applied to accelerate the procedures \citep{VO:2007}.
BWD is also easy to implement and of complexity $O(n\log n)$ \citep{Shin2014}.

\subsection{$\ell_1$-penalization}\label{S3.3}
We discussed in Section \ref{regmod} that detecting change points is essentially a regression problem with sparsity. Therefore, it is not surprising that there have been attempts to use the methods from penalized regression to the detection of change points.
\cite{HuangEtal:2005,harchaoui2010multiple} studied the following optimization problem
\begin{eqnarray}\label{l1a}
\text{minimize } ||\bY-\btheta||^2 \qquad \text{subject to }  \quad \sum_{j=1}^{n-1}|\theta_j-\theta_{j+1}|\leq s.
\end{eqnarray}
Through reparametrization $\beta_j=\theta_{j+1}-\theta_{j}$, the above optimization problem is equivalent to
\begin{eqnarray}\label{l1b}
\text{minimize } ||\bY-\bX\bbeta||^2 \qquad \text{subject to }  \quad \sum_{j=1}^{n-1}|\beta_j|\leq s,
\end{eqnarray}
which gives the LASSO solution for regression model (\ref{aaa4}). (\ref{l1a}) and (\ref{l1b}) are usually solved through their dual forms

\begin{eqnarray*}
 \text{minimize }  ||\bY-\btheta||^2 +\lambda \sum_{j=1}^{n-1}|\theta_j-\theta_{j+1}| \quad \text{ or }\quad  ||\bY-\bX\bbeta||^2 + \lambda \sum_{j=1}^{n-1}|\beta_j|,
\end{eqnarray*}
where $\lambda \sum_{j=1}^{n-1}|\theta_j-\theta_{j+1}|$ is called the total variation penalty, $\lambda \sum_{j=1}^{n-1}|\beta_j|$ is called $\ell_1$ penalty or LASSO penalty.

Furthermore, \cite{fusedlasso:2008} applied the fused LASSO \citep{Tibshirani:fusedlasso:2005} for change-point detection. In particular, they estimated $\btheta$ through the following constrained optimization:
\begin{eqnarray}\label{flasso}
\quad \text{ minimize } ||\bY-\btheta||^2 \qquad \text{subject to } ||\btheta||_{\ell_1}\leq s_1,\quad \sum_j|\theta_j-\theta_{j+1}|\leq s_2,
\end{eqnarray}
or equivalently,
\[\text{minimize } ||\bY-\bX\bbeta||^2 \qquad \text{subject to } \sum_{j=1}^n\left|\sum_{k=0}^{j-1}\beta_k\right|\leq s_1,\quad \sum_{j=1}^{n-1}|\beta_j|\leq s_2.\]
The difference between (\ref{l1a}) and (\ref{flasso}) lies in whether $\btheta$ is also assumed to be sparse. In the context of detecting chromosome copy number variations, it is reasonable to assume that $\btheta$ is sparse because of the discrete nature of the copy numbers. From the methodological and computational perspectives, the solution for (\ref{flasso}) can be obtained by simply thresholding the solution of (\ref{l1a}) \citep{CDA:2007}. Hence, there is little loss of generality by focusing on (\ref{l1a}) and (\ref{l1b}).

For the $\ell_1$ penalization methods, standard regression solvers such as the LARS \citep{LARS:2004} and coordinate decent algorithm \citep{CDA:2007} can be used to solve optimization problem (\ref{l1b}). The complexity of some related change-point detection algorithms can achieve $O(K_{\max}n\log n)$, where $K_{\max}$ is an upper bound for the number of change points \citep{harchaoui2010multiple}.

\blue{It is crucial to determine the values of tuning parameters in these procedures. Some tuning parameter selection methods have been used in the literature. In particular, \cite{HuangEtal:2005} proposed to choose $s$ empirically. They examined the solutions of (20) for an increasing sequence of $s$. As more change points are added into the model with larger values of $s$, they chose to stop increasing $s$ when the resulted mean difference at the new change point is not big enough. \cite{fusedlasso:2008} estimated $s_1$ and $s_2$ for (22) based on heavily and moderately smoothed versions of $\bY$, respectively. \cite{harchaoui2010multiple} gave an asymptotic order of the tuning parameter $\lambda$ in the dual optimization problem. \cite{CDA:2007} recommended cross-validation to select tuning parameters for the fused LASSO. As stated in \cite{Rinaldo:2009}, it is an important open problem to find the optimal values for the parameters.} 

\subsection{Screening and ranking algorithm}\label{S3.4}
A Screening and Ranking algorithm (SaRa) has been studied in \cite{NiuZhang:2010,HaoNiuZhang2013} to detect change points. For the normal mean model (\ref{aaa2}), they considered the locally defined statistic at each position $h\leq j\leq n-h$,
 \[D_h(j)= \left(\sum_{k=j+1}^{j+h} Y_k-\sum_{k=j-h+1}^j Y_k\right)/h ,\]
where $h$ is a fixed bandwidth. Essentially, $D_h(j)$ is the likelihood ratio test statistic for the local testing problem (\ref{aaa8}). Hence, the sequence \{$D_h(\cdot)$\} roughly measures the relative likelihood for each position to be a change point.
The SaRa proceeds as follows. First, it calculates $D_h(\cdot)$ and finds all local maximizers of $|D_h(\cdot)|$. Here, $j$ is a local maximizer if $|D_h(j)|\geq |D_h(k)|$ for all $k\in (j-h,j+h)$. Second, the SaRa estimator is obtained by applying a thresholding rule $|D_h(\cdot)|>\lambda$ to all local maximizers. Consequently, the estimated change point locations are a set of ordered positions
\[\mathcal{J}_{h,\lambda}=\{\hat\tau_{\kappa}:\hat\tau_{\kappa} \text{ is a local maximizer of } |D_h(\cdot)|, \text{ and } |D_h(\hat\tau_{\kappa})|>\lambda\}.\]
Let $\hat{J}=|\mathcal{J}_{h,\lambda}|$. We let $\hat{\btau}=(\hat\tau_1,...,\hat\tau_{\hat{J}})^{\top}$ denote the SaRa estimator.

In fact, the SaRa procedure can be described as a modified version of binary segmentation. At the first step, $|D_h(\cdot)|$ is calculated and its global maximizer divides the sequence into two segments. Then in each segment, we can repeat the same procedure until the maximum of $|D_h(\cdot)|$ is below some threshold for each segment.
Therefore, the difference between the SaRa and BS is the use of a local or global likelihood ratio test statistic as the basis of optimization. The SaRa has several advantages. First, the test statistic does not have to be recalculated every step. Second, one does not have to conduct the SaRa stagewisely, and a unified threshold can be applied simultaneously. Moreover, with the framework of the SaRa, it is possible to solve change-point problem as a multiple testing one (\ref{aaa8}).

The computational complexity of the SaRa is only at $O(n)$ because the sequence $D_h(\cdot)$ can be calculated by a \red{recursive} formula \citep{NiuZhang:2010}.

\subsection{SMUCE}\label{S3.5}
\cite{frick2014multiscale} introduced a new change-point detection tool called simultaneous multiscale change-point estimator (SMUCE). It estimates the mean vector $\btheta$ by solving an optimization problem
\begin{eqnarray}\label{aaa24}
\text{minimize } |J(\btheta)| \text{ subject to } T_n(\bY,\btheta)\leq q,
\end{eqnarray}
where $|J(\btheta)|$ is the number of change points along piecewise constant vector $\btheta$, $T_n(\bY,\btheta)$ is a multiscale statistic defined below and $q$ is a threshold. For a fixed $\btheta$,
\begin{eqnarray}\label{multiscale}
\quad T_n(\bY,\btheta)= \max_{1\leq i\leq j\leq n;\theta_i=\theta_{i+1}=\cdots=\theta_{j}} \left[\sqrt{2T_i^j(\bY,\theta_i)}-\sqrt{2\log\frac{en}{j-i+1}}\right],
\end{eqnarray}
where $T_i^j(\bY,\theta_i)$ is a local log-likelihood ratio test statistic for testing $H_0:\theta=\theta_i$ versus $H_0:\theta\ne\theta_i$ on the interval $[i,j]$. In the context of normal mean MCM,
\[T_i^j(\bY,\theta_i)=\frac{j-i+1}{2}(\bar Y_{i}^j-\theta_i)^2,\]
where $\bar Y_{i}^j=\frac{1}{j-i+1}\sum_{\ell=i}^jY_{\ell}$. Note that $T_i^j$ is defined only on the interval $[i,j]$ where $\btheta$ is constant and it reflects the local discrepancy of the model $\btheta$ and the data.
That is, $T_n(\bY,\btheta)$ is \blue{an} aggregation of local \red{statistics} $T_i^j$. Nevertheless, (\ref{aaa24}) is a global optimization problem.
\cite{frick2014multiscale} employs dynamic programming technique to solve SMUCE. The general complexity is $O(n^2)$, and may be reduced under certain conditions.
\blue{In a recent work, \cite{pein2015heterogeneuous} further extended SMUCE to the heterogeneous case.}

\section{Consistency, convergence rate and simultaneous inference}\label{S5}
In this section, we discuss theoretical results on change-point analysis. Recall that for normal mean MCM (\ref{aaa2}), we define $L=\min_{0\leq \kappa\leq J}\{\tau_{\kappa+1}-\tau_{\kappa}\}$, $\delta=\min_{1\leq i<n}\{|\theta_i-\theta_{i+1}|\}/\sigma$, and $S=\delta^2L$ which represents the overall signal strength. For asymptotic theory, we typically assume that all model parameters and these important quantities depends on $n$ and study the asymptotic behavior of an estimator $\hat\btau$ as $n\to\infty$.

\subsection{Simple cases}\label{S5.1}
Models (\ref{aaa11}) and (\ref{aaa15}) are the two simplest and best understood change-point models. We start with the single change-point model (\ref{aaa11}), studied in, e.g.,  \cite{hinkley1970inference,CsorgoHorvath:1997}. If $H_1$ is true, and the change-point location ${\tau}$ and mean jump at the change point $\delta=|\theta_{{\tau}+1}-\theta_{\tau}|/\sigma$ satisfy either of the following two conditions
\begin{eqnarray}
0<\frac{{\tau}}{n}\to t<1, \quad \delta\to 0, \text{ with } \lim_{n\to \infty}\frac{n\delta^2}{\log\log n}=\infty;\label{asymp1}\\
\frac{{\tau}}{n}\to 0,\quad \delta\to 0, \text{ with } \lim_{n\to \infty}\frac{{\tau}\delta^2}{\log\log n}=\infty,\label{asymp2}
\end{eqnarray}
then the maximum likelihood estimator $\hat{\tau}=\hat j$ in (\ref{aaa14}) satisfies
\begin{eqnarray}\label{consistency2}
\delta^2|\hat{{\tau}}-{\tau}|=O_P(1).
\end{eqnarray}
For a less challenging setting  $\delta \to c>0$, it implies
\begin{eqnarray}\label{aaa30}
|\hat{{\tau}}-{\tau}|=O_P(1).
\end{eqnarray}
It was regarded as a result of inconsistency in \cite{hinkley1970inference} because there is no way to recover exact change-point location with an overwhelming probability. But at the same time, it was interpreted as a consistency result in the literature since it indicates
\begin{eqnarray}\label{consistency3}
\left|\frac{\hat {\tau}}{n}-\frac{{\tau}}{n}\right|=O_P\left(\frac1n\right).
\end{eqnarray}
That is, if we embed the sequence into the unit interval, the change-point location $\frac{{\tau}}{n}$ can be estimated consistently with the convergence rate $\frac1n$. This is an optimal result and the conditions (\ref{asymp1}) and (\ref{asymp2}) can not be relaxed further \citep{CsorgoHorvath:1997}.

\cite{Donoho:Huo:2005} studied the epidemic change-point model (\ref{aaa15}) and concluded that no method can detect the mean shift reliably when $\delta^2(r-l) <2\log n$. Moreover, they proposed a near-optimal procedure which can efficiently and reliably detect the mean shift segment if $\delta^2(r-l)$ is slightly above $2\log n$. Their work suggests that $2\log n$ be the optimal detection threshold for (\ref{aaa15}) and offers a benchmark for the necessary conditions to solve the general change-point problem.

\subsection{General cases}\label{S5.2}
For general cases, the consistency result includes two parts, on the number $J^*$ and location $\btau^*$ of change points. That is,
\begin{eqnarray}
\lim_{n\to\infty}P(\hat J=J^*)&=&1,\\
\delta^2\|\hat\btau-\btau^*\|_{\infty}&=&O_P(1).
\end{eqnarray}
Such results have been obtained by \cite{Yao1988,Yao:1989} for the exhaustive search method under an asymptotic setting that $J^*$ and $\delta$ are fixed, and $\btau^*/n\to \bt$ as $n\to \infty$, where $\bt=(t_1,\cdots,t_{J^*})^{\top}$ with $0<t_1<\cdots<t_J<1$ is a constant vector. This asymptotic framework corresponds to category S1 listed in Table \ref{Tb:signal} of Section \ref{S2.4}.

Using regression approach and the LASSO techniques, \cite{Rinaldo:2009,qian2012pattern} obtained consistency result $\bP(\hat\btau=\btau^*)\to 1$. However, a strong condition $\delta^2\gg\log n$ is typically required in most of the related theory with an exception of \cite{harchaoui2010multiple}. We listed this framework as category S2 in Table \ref{Tb:signal}. The result is less useful because it relies only on big mean changes and not on the distance between the change points.

We should note that a naive approach can also achieve the consistency.
\begin{proposition}\label{P1}
Let $z_j=Y_{j+1}-Y_{j}$ for $j=1,...,n-1$. Consider a naive estimator $\hat\btau_{naive}(c)$ that is the vector of ordered elements in set $\{j: |z_j|>c, 1\leq j\leq n-1\}$. With $\delta=\min_{j\in\btau^*}|\theta_{j+1}-\theta_j|/\sigma>4\sqrt{ \log n}$, we have $\bP(\hat\btau_{naive}(c)=\btau^*)\to 1$ for $c=2\sigma\sqrt{\log n}$.
\end{proposition}

{\bf Proof.} Consider two events $\cA=\{|z_j|>c \text{ for all } j\in \btau^*\}$ and $\cB=\{|z_j|<c \text{ for all } j\notin \btau^*\}$. It is sufficient to show $\bP(\cA\cap\cB)\to 1$ as $n\to \infty$. By definition, we have $z_j\sim\cN(\theta_{j+1}-\theta_j,2\sigma^2)$ for all $1\leq j\leq n-1$. Define $\tilde z_j=\left(z_j-(\theta_{j+1}-\theta_j)\right)/(\sqrt{2}\sigma)\sim\cN(0,1)$. Then
\[\bP\left(\max_{1\leq j\leq n-1} \{\tilde z_j^2\}< 2\log n\right)\to 1,\quad n\to \infty,\]
which implies $\bP(\cA\cap\cB)\to 1$, provided $c=2\sigma\sqrt{\log n}$ and $\delta>4\sqrt{ \log n}$. It is easy to see that the result holds for $\delta>2(\sqrt{ \log n}+\sqrt{\log J^*})$ if the number of change points $J^*\ll n$. $\square$

Following the work reviewed in Section \ref{S5.1}, there is a series of \red{results that} relax the conditions such as $S=\delta^2L>C\log n$ or $S\gg \log n$. In particular, \cite{harchaoui2010multiple} studied $\ell_1$ penalization approach and showed consistency result with a rate slightly slower than $O_P(\log n/\delta^2)$ for change-point locations, under the condition $\delta^2L\gg\log n$. \cite{HaoNiuZhang2013} obtained optimal $O_P(1/\delta^2)$ convergence with condition $\delta^2L>32\log n$ for the SaRa. \cite{fryzlewicz2012wild} proposed WBS which requires $\delta^2L\gg C\log n$ for sufficiently large $C$ to achieve convergence rate $O_P(\log n/\delta^2)$. SMUCE also requires $\delta^2L\gg\log n$ \citep{frick2014multiscale}. In spite of these results, the \red{optimal} condition to achieve consistency remains to be an open issue.

We presented above sufficient conditions for different methods to obtain consistency result. However, the necessary conditions are \red{rarely} discussed in the literature. Loosely speaking, $S\geq 2\log n$ appears necessary as we discussed in Section \ref{S5.1}. Nonetheless, \red{finding necessary conditions to assure consistency is an important line of inquiry.}

\subsection{\blue{Simultaneous inference}}

\blue{\cite{hinkley1970inference,worsley1986confidence,siegmund1988confidence} studied confidence interval construction for single change point models. For MCM, it is natural to ask how to
\begin{enumerate}
\item construct simultaneous confidence intervals for change-point locations and
\item assign significance simultaneously for all detected change points.
\end{enumerate}
\red{These two questions are distinct but related}. The first one concerns} \red{with assessing} \blue{the accuracy and uncertainty of a point estimator for change-point location. When the signal strength is relatively weak, it may not be practical to recover and construct confidence intervals for all change points. \red{However}, it is desirable to assign significance} \red{level} \blue{for each detected change point, which will help to identify as many change points as possible while controlling \red{for} false positives. }

\blue{As far as we know, not much work has been done on these important topics until very recently. \cite{frick2014multiscale} proposed SMUCE} to construct confidence intervals for multiple change points as well as the stepwise mean function. \blue{In addition,} the familywise error rate (FWER) for estimated change points can be controlled through a tuning parameter. 
\blue{\cite{HaoNiuZhang2013} discussed the second question and established a framework to control FDR via the SaRa. In particular, they considered} the multiple testing formulation introduced in Section \ref{S2.2.3}. \blue{A key point is that, \red{for each change point detected by the SaRa that is a false positive,} the distribution of $p$ value for the local test (\ref{aaa8}) can be obtained. Hence, the significance levels of all detected change points can be assigned by a simple transformation of the local test statistic. Moreover, the detected change-point locations are well separated so the FDR can be} \red{assessed} \blue{easily by standard Benjamini-Hochberg procedure \citep{BH:95}.}

\blue{For any error control procedure, one important and subtle point is to define ``true positive'' for a detected change-point location. In a standard multiple testing framework without a sequential structure among the tests, it is straightforward to define a true positive. However, \red{this is overly restrictive for change-point problems.} Recall that in (\ref{aaa6})-(\ref{aaa8}) $H_1(j)$ is true when $j$ is a change point. In the classical results such as the one presented in (\ref{aaa30}), the detected change point location is not expected to be exactly the same as the true one with overwhelming probability. That is, it is very likely that $H_0(j)$ is rejected \red{when the position $j$ is close }to a true change point. Obviously, it is reasonable to treat it as true positive if $H_0(j)$ is the only rejected null hypothesis in a small neighborhood of a true change-point location $\tau$. This strategy was used in \cite{HaoNiuZhang2013}. In particular, they treated an estimate $\hat\tau$ as revealing a true change point $\tau$ if they are close enough, say, $|\hat\tau-\tau|<h$, where $h$ is the same \red{window size} as in (\ref{aaa8}).}

\red{Except SaRa and SMUCE, the methods for MCM have not been well understood theoretically, and it remains to establish an inferential framework for those methods.} 

\section{Aggregating local information}\label{S6}
For some of aforementioned algorithms including exhaustive search, dynamic programming, $\ell_1$-penalization, BS and CBS, all the data are involved directly and simultaneously in an optimization procedure or the initial step of a sequential test procedure. We may call them \emph{global} methods, which have been playing a dominant role in change-point analysis. On the other hand, a variety of \emph{local} methods have been developed recently. Those local methods share a common strategy, i.e., to first gather and then aggregate local information. Among the methods reviewed in this paper, we may consider the SaRa, WBS and BWD as examples of local methods. In addition, SMUCE also shares \red{a} similar spirit.

Change-point inference is often stated \red{in} a hypothesis testing \red{framework} of the form (\ref{aaa5}). Exhaustive search aims to solve (\ref{aaa5}) directly by searching for the strongest evidence against the null hypothesis over all possible location combinations. Forward stepwise methods BS and CBS solve testing problems of types (\ref{aaa11}) and (\ref{aaa15}) recursively. In contrast, local methods start from various localized versions of (\ref{aaa11}). For example, the SaRa focuses on a sequence of local tests (\ref{aaa8}), and makes decision based on local maxima of the test statistic sequence and a global thresholding rule; WBS considers test (\ref{testSE}) defined on a random interval $[s,e]$, and then combines the test statistics over a set of random intervals; BWD considers a sequence of local test problems (\ref{aaa19}) \red{ at each step, and retain a single} null hypothesis by comparing local test statistics. We see that the SaRa, WBS and BWD focus first tests on small local segments then make decisions based on \blue{an} aggregation of the local test statistics. Note that for BS and CBS, the alternative hypothesis is never truly correct \red{except} the very last step. Therefore, the \red{signal} brought by \red{each change point may be weakened by other change points, leading to reduced power.} \blue{To elaborate, consider a sequence of Gaussian random variables of length $n=1000$ with mean $\theta_i=1$ when $500<i\leq 510$ and 0 otherwise.} \red{When BS is applied,} \blue{the test statistic (\ref{aaa12}) may not be significant as the signals brought by the two change points are neutralized. Thus, aggregating local tests} should avoid signal cancellation caused by multiple change points. As illustrated in Section \ref{S2.3}, a good method may mimic the oracle to perform single change-point test on small segments, each of which has at most one change point. To this end, WBS draws a set of random segments in each step and hopes that some of these segments are close to the ones in the oracle division; SaRa checks a small neighborhood of each position; BWD builds up segments \red{adaptively from the data beginning with single point segments}. SMUCE estimates change points by solving a global optimization problem. Nevertheless, the employed multiscale statistic (\ref{multiscale}) can be considered as an aggregation of local statistics.

\blue{In spite of these recent developments, researchers are still looking for better methods to aggregate local information, especially, when little information on the distances among change points} \red{is} \blue{available. One difficulty is how to determine a local neighborhood when calculating a local statistic. Intuitively, it may involve two or more change points if \red{too} large neighborhood is used. On the other hand, it} \red{may reduce the} \blue{power if a small neighborhood is used. Moreover, it \red{may not be} straightforward to compare and aggregate local statistics if they are calculated based on neighborhoods \red{having} different sizes.} 

\section{Concluding remarks}\label{S7}
In spite of recent rapid developments on this topic, there are \red{many} interesting and open research directions to explore such as optimal detection and simultaneous inference for multiple change points. In contrast to classical approaches which solve global optimization to estimate change points, there are emerging new methods which decompose MCM into a set of local problems, and then gather and aggregate local information to solve MCM. There are several advantages \red{to} this strategy. First, the local information is usually summarized by \red{a single} statistic for each local problem. \blue{A typical approach} to \red{aggregating} these statistics \blue{is to find} (local or global) extreme values. Therefore, the computation is less expensive as no complex optimization is involved. Second, these local statistics together bring more information than a point estimator so inference is possible. For example, \blue{in the SaRa, local maxima are extracted to represent the likelihood} \red{for the} \blue{presence of change points, and the local statistics around local maxima are ignored but potentially useful for inference. So far,} \red{\blue{besides recent works} \citep{HaoNiuZhang2013,frick2014multiscale}, a solid framework for inference on multiple change points based on local methods is lacking}. It is important to find the best way to gather and aggregate the local information.


\red{Without prior knowledge, a change point may be located anywhere along the sequence.} It is well known that there is no optimal test for (\ref{aaa1}) because of the indeterminacy of the change-point location \citep{SenSri:1975}. We may have one procedure that is more powerful when the change point is in the middle and the other one more powerful when it is close to the boundary. As illustrated in Section \ref{S2.3}, we may consider the circular model to make the change-point problem symmetric. The symmetry comes from the finite group action $\mathbb{Z}_n$ to the change-point locations, where $\mathbb{Z}_n$ is a cyclic group of order $n$. Under this new formulation, we may study equivariant change-point detection procedures. Among the aforementioned algorithms, CBS is designed to solve the circular model, while exhaustive search, SMUCE, SaRa and BWD are able to solve the circular model after slight modifications. \red{Studying theoretical properties of equivariance procedures for the symmetric model remains an open topic}.  

It seems that two striking features of MCM, i.e., locality and symmetry, have not been emphasized in the literature. We hope that this paper can stimulate more research on these aspects.


\section*{Acknowledgements}
This research was partially supported by grants DMS 1309507 from National Science Foundation and R01 DA016750 from the National Institutes of Health. \blue{The authors are grateful to Joseph Watkins for helpful comments.}

\bibliographystyle{biometrika}
\bibliography{changepointrefs2014}

\begin{thebibliography}{46}
\expandafter\ifx\csname natexlab\endcsname\relax\def\natexlab#1{#1}\fi

\bibitem[{Arias-Castro et~al.(2005)Arias-Castro, Donoho \&
  Huo}]{Donoho:Huo:2005}
\textsc{Arias-Castro, E.}, \textsc{Donoho, D.~L.} \& \textsc{Huo, X.} (2005).
\newblock Near-optimal detection of geometric objects by fast multiscale
  methods.
\newblock \textit{IEEE Transactions on Information Theory} \textbf{51},
  2402--2425.

\bibitem[{Benjamini \& Hochberg(1995)}]{BH:95}
\textsc{Benjamini, Y.} \& \textsc{Hochberg, Y.} (1995).
\newblock {Controlling the False Discovery Rate: A Practical and Powerful
  Approach to Multiple Testing}.
\newblock \textit{{Journal of the Royal Statistical Society. Series B }}
  \textbf{57}, 289--300.

\bibitem[{Braun et~al.(2000)Braun, Braun \& Muller}]{BBM:2000}
\textsc{Braun, J.~V.}, \textsc{Braun, R.~K.} \& \textsc{Muller, H.~G.} (2000).
\newblock Multiple changepoint fitting via quasilikelihood, with application to
  {DNA} sequence segmentation.
\newblock \textit{Biometrika} \textbf{87}, 301--314.

\bibitem[{Brodsky \& Darkhovsky(2000)}]{brodsky2000non}
\textsc{Brodsky, B.} \& \textsc{Darkhovsky, B.} (2000).
\newblock \textit{Non-Parametric Statistical Diagnosis: Problems and Methods}.
\newblock Mathematics and Its Applications. Springer.

\bibitem[{Brodsky \& Darkhovsky(1993)}]{brodsky1993nonparametric}
\textsc{Brodsky, E.} \& \textsc{Darkhovsky, B.} (1993).
\newblock \textit{Nonparametric Methods in Change Point Problems}.
\newblock Mathematics and Its Applications. Springer.

\bibitem[{Carlin et~al.(1992)Carlin, Gelfand \& Smith}]{carlin1992hierarchical}
\textsc{Carlin, B.~P.}, \textsc{Gelfand, A.~E.} \& \textsc{Smith, A.~F.}
  (1992).
\newblock Hierarchical bayesian analysis of changepoint problems.
\newblock \textit{Applied statistics} , 389--405.

\bibitem[{Carlstein et~al.(1994)Carlstein, M{\"u}ller \&
  Siegmund}]{carlstein1994change}
\textsc{Carlstein, E.}, \textsc{M{\"u}ller, H.} \& \textsc{Siegmund, D.}
  (1994).
\newblock \textit{Change-point Problems}.
\newblock No. v. 23 in Change-Point Problems. Institute of Mathematical
  Statistics.

\bibitem[{Chen \& Gupta(2000)}]{ChenGupta:2000:book}
\textsc{Chen, J.} \& \textsc{Gupta, A.} (2000).
\newblock \textit{Parametric statistical change point analysis}.
\newblock Dmv Seminar Series. Birkh{\"a}user.

\bibitem[{Cobb(1978)}]{cobb1978problem}
\textsc{Cobb, G.~W.} (1978).
\newblock The problem of the nile: conditional solution to a changepoint
  problem.
\newblock \textit{Biometrika} \textbf{65}, 243--251.

\bibitem[{Cs{\"o}rg{\"o} \& Horv{\'a}th(1997)}]{CsorgoHorvath:1997}
\textsc{Cs{\"o}rg{\"o}, M.} \& \textsc{Horv{\'a}th, L.} (1997).
\newblock \textit{Limit theorems in change-point analysis}.
\newblock Wiley series in probability and statistics. Wiley.

\bibitem[{Efron et~al.(2004)Efron, Hastie, Johnstone \& Tibshirani}]{LARS:2004}
\textsc{Efron, B.}, \textsc{Hastie, T.}, \textsc{Johnstone, I.} \&
  \textsc{Tibshirani, R.} (2004).
\newblock Least angle regression.
\newblock \textit{Ann. Stat.} \textbf{32}, 407--499.

\bibitem[{Frick et~al.(2014)Frick, Munk \& Sieling}]{frick2014multiscale}
\textsc{Frick, K.}, \textsc{Munk, A.} \& \textsc{Sieling, H.} (2014).
\newblock Multiscale change point inference.
\newblock \textit{Journal of the Royal Statistical Society: Series B
  (Statistical Methodology)} \textbf{76}, 495--580.

\bibitem[{Friedman et~al.(2007)Friedman, Hastie, H\"{o}fling \&
  Tibshirani}]{CDA:2007}
\textsc{Friedman, J.}, \textsc{Hastie, T.}, \textsc{H\"{o}fling, H.} \&
  \textsc{Tibshirani, R.} (2007).
\newblock Pathwise coordinate optimization.
\newblock \textit{Ann. Appl. Stat.} \textbf{1}, 302--332.

\bibitem[{Fryzlewicz et~al.(2014)}]{fryzlewicz2012wild}
\textsc{Fryzlewicz, P.} et~al. (2014).
\newblock Wild binary segmentation for multiple change-point detection.
\newblock \textit{The Annals of Statistics} \textbf{42}, 2243--2281.

\bibitem[{Hao et~al.(2013)Hao, Niu \& Zhang}]{HaoNiuZhang2013}
\textsc{Hao, N.}, \textsc{Niu, Y.~S.} \& \textsc{Zhang, H.} (2013).
\newblock Multiple change-point detection via a screening and ranking
  algorithm.
\newblock \textit{Statistica Sinica} \textbf{23}, 1553--1572.

\bibitem[{Harchaoui \& L{\'e}vy-Leduc(2010)}]{harchaoui2010multiple}
\textsc{Harchaoui, Z.} \& \textsc{L{\'e}vy-Leduc, C.} (2010).
\newblock Multiple change-point estimation with a total variation penalty.
\newblock \textit{Journal of the American Statistical Association}
  \textbf{105}, 1480--1493.

\bibitem[{Hawkins(1977)}]{Hawkins:1977}
\textsc{Hawkins, D.~M.} (1977).
\newblock Testing a sequence of observations for a shift in location.
\newblock \textit{Journal of the American Statistical Association} \textbf{72},
  pp. 180--186.

\bibitem[{Hinkley(1970)}]{hinkley1970inference}
\textsc{Hinkley, D.~V.} (1970).
\newblock Inference about the change-point in a sequence of random variables.
\newblock \textit{Biometrika} \textbf{57}, 1--17.

\bibitem[{Huang et~al.(2005)Huang, Wu, Lizardi \& Zhao}]{HuangEtal:2005}
\textsc{Huang, T.}, \textsc{Wu, B.}, \textsc{Lizardi, P.} \& \textsc{Zhao, H.}
  (2005).
\newblock Detection of {DNA} copy number alterations using penalized least
  squares regression.
\newblock \textit{Bioinformatics} \textbf{21}, 3811--3817.

\bibitem[{Jackson et~al.(2005)Jackson, Scargle, Barnes, Arabhi, Alt,
  Gioumousis, Gwin, Sangtrakulcharoen, Tan \& Tsai}]{jackson2005algorithm}
\textsc{Jackson, B.}, \textsc{Scargle, J.~D.}, \textsc{Barnes, D.},
  \textsc{Arabhi, S.}, \textsc{Alt, A.}, \textsc{Gioumousis, P.}, \textsc{Gwin,
  E.}, \textsc{Sangtrakulcharoen, P.}, \textsc{Tan, L.} \& \textsc{Tsai, T.~T.}
  (2005).
\newblock An algorithm for optimal partitioning of data on an interval.
\newblock \textit{Signal Processing Letters, IEEE} \textbf{12}, 105--108.

\bibitem[{Killick et~al.(2012)Killick, Fearnhead \&
  Eckley}]{killick2012optimal}
\textsc{Killick, R.}, \textsc{Fearnhead, P.} \& \textsc{Eckley, I.} (2012).
\newblock Optimal detection of changepoints with a linear computational cost.
\newblock \textit{Journal of the American Statistical Association}
  \textbf{107}, 1590--1598.

\bibitem[{Lee(2010)}]{lee2010change}
\textsc{Lee, T.-S.} (2010).
\newblock Change-point problems: bibliography and review.
\newblock \textit{Journal of Statistical Theory and Practice} \textbf{4},
  643--662.

\bibitem[{Levin \& Kline(1985)}]{levin1985cusum}
\textsc{Levin, B.} \& \textsc{Kline, J.} (1985).
\newblock The cusum test of homogeneity with an application in spontaneous
  abortion epidemiology.
\newblock \textit{Statistics in Medicine} \textbf{4}, 469--488.

\bibitem[{Niu \& Zhang(2012)}]{NiuZhang:2010}
\textsc{Niu, Y.~S.} \& \textsc{Zhang, H.} (2012).
\newblock The screening and ranking algorithm to detect {DNA} copy number
  variations.
\newblock \textit{The Annals of Applied Statistics} \textbf{6}, 1306--1326.

\bibitem[{Olshen et~al.(2004)Olshen, Venkatraman, Lucito \& Wigler}]{OVLW:2004}
\textsc{Olshen, A.~B.}, \textsc{Venkatraman, E.~S.}, \textsc{Lucito, R.} \&
  \textsc{Wigler, M.} (2004).
\newblock Circular binary segmentation for the analysis of array-based {DNA}
  copy number data.
\newblock \textit{Biostatistics} \textbf{5}, 557--72.

\bibitem[{Page(1954)}]{page1954continuous}
\textsc{Page, E.} (1954).
\newblock Continuous inspection schemes.
\newblock \textit{Biometrika} \textbf{41}, 100--115.

\bibitem[{Page(1955)}]{Page:1955}
\textsc{Page, E.~S.} (1955).
\newblock A test for a change in a parameter occurring at an unknown point.
\newblock \textit{Biometrika} \textbf{42}, pp. 523--527.

\bibitem[{Page(1957)}]{Page:1957}
\textsc{Page, E.~S.} (1957).
\newblock On problems in which a change in a parameter occurs at an unknown
  point.
\newblock \textit{Biometrika} \textbf{44}, pp. 248--252.

\bibitem[{Pein et~al.(2015)Pein, Sieling \& Munk}]{pein2015heterogeneuous}
\textsc{Pein, F.}, \textsc{Sieling, H.} \& \textsc{Munk, A.} (2015).
\newblock Heterogeneuous change point inference.
\newblock \textit{arXiv preprint arXiv:1505.04898} .

\bibitem[{Qian \& Jia(2012)}]{qian2012pattern}
\textsc{Qian, J.} \& \textsc{Jia, J.} (2012).
\newblock On pattern recovery of the fused lasso.
\newblock \textit{arXiv preprint arXiv:1211.5194} .

\bibitem[{Rinaldo(2009)}]{Rinaldo:2009}
\textsc{Rinaldo, A.} (2009).
\newblock Properties and refinements of the fused lasso.
\newblock \textit{Ann. Statist.} \textbf{37}, 2922--2952.

\bibitem[{Scott \& Knott(1974)}]{scott1974cluster}
\textsc{Scott, A.} \& \textsc{Knott, M.} (1974).
\newblock A cluster analysis method for grouping means in the analysis of
  variance.
\newblock \textit{Biometrics} , 507--512.

\bibitem[{Sen \& Srivastava(1975)}]{SenSri:1975}
\textsc{Sen, A.} \& \textsc{Srivastava, M.~S.} (1975).
\newblock On tests for detecting change in mean.
\newblock \textit{The Annals of Statistics} \textbf{3}, pp. 98--108.

\bibitem[{Shin et~al.(2014)Shin, Wu \& Hao}]{Shin2014}
\textsc{Shin, S.~J.}, \textsc{Wu, Y.} \& \textsc{Hao, N.} (2014).
\newblock A backward procedure for change-point detection with applications to
  copy number variation detection.
\newblock \textit{manuscript} .

\bibitem[{Siegmund(1988)}]{siegmund1988confidence}
\textsc{Siegmund, D.} (1988).
\newblock Confidence sets in change-point problems.
\newblock \textit{International Statistical Review/Revue Internationale de
  Statistique} , 31--48.

\bibitem[{Tibshirani et~al.(2005)Tibshirani, Saunders, Rosset, Zhu \&
  Knight}]{Tibshirani:fusedlasso:2005}
\textsc{Tibshirani, R.}, \textsc{Saunders, M.}, \textsc{Rosset, S.},
  \textsc{Zhu, J.} \& \textsc{Knight, K.} (2005).
\newblock Sparsity and smoothness via the fused lasso.
\newblock \textit{Journal Of The Royal Statistical Society Series B}
  \textbf{67}, 91--108.

\bibitem[{Tibshirani \& Wang(2008)}]{fusedlasso:2008}
\textsc{Tibshirani, R.} \& \textsc{Wang, P.} (2008).
\newblock Spatial smoothing and hot spot detection for cgh data using the fused
  lasso.
\newblock \textit{Biostatistics (Oxford, England)} \textbf{9}, 18--29.

\bibitem[{Venkatraman \& Olshen(2007)}]{VO:2007}
\textsc{Venkatraman, E.~S.} \& \textsc{Olshen, A.~B.} (2007).
\newblock A faster circular binary segmentation algorithm for the analysis of
  array {CGH} data.
\newblock \textit{Bioinformatics} \textbf{23}, 657--663.

\bibitem[{Vostrikova(1981)}]{Vostrikova:1981}
\textsc{Vostrikova, L.~Y.} (1981).
\newblock Detecting ``disorder'' in multidimensional random processes.
\newblock \textit{Soviet Mathematics Doklady} , pp.55--59.

\bibitem[{Worsley(1986)}]{worsley1986confidence}
\textsc{Worsley, K.~J.} (1986).
\newblock Confidence regions and tests for a change-point in a sequence of
  exponential family random variables.
\newblock \textit{Biometrika} \textbf{73}, 91--104.

\bibitem[{Yao(1993)}]{YaoQ:1993}
\textsc{Yao, Q.} (1993).
\newblock Tests for change-points with epidemic alternatives.
\newblock \textit{Biometrika} \textbf{80}, pp. 179--191.

\bibitem[{Yao(1988)}]{Yao1988}
\textsc{Yao, Y.-C.} (1988).
\newblock Estimating the number of change-points via schwarz' criterion.
\newblock \textit{Statistics \& Probability Letters} \textbf{6}, 181--189.

\bibitem[{Yao \& Au(1989)}]{Yao:1989}
\textsc{Yao, Y.-C.} \& \textsc{Au, S.~T.} (1989).
\newblock Least-squares estimation of a step function.
\newblock \textit{Sankhya-: The Indian Journal of Statistics, Series A}
  \textbf{51}, 370--381.

\bibitem[{Yin(1988)}]{Yin88}
\textsc{Yin, Y.~Q.} (1988).
\newblock Detection of the number, locations and magnitudes of jumps.
\newblock \textit{Communications in Statistics. Stochastic Models} \textbf{4},
  445 -- 455.

\bibitem[{Zhang \& Siegmund(2007)}]{ZhangSiegmund:2007}
\textsc{Zhang, N.~R.} \& \textsc{Siegmund, D.~O.} (2007).
\newblock A modified bayes information criterion with applications to the
  analysis of comparative genomic hybridization data.
\newblock \textit{Biometrics} \textbf{63}, 22--32.

\bibitem[{Zhang et~al.(2010)Zhang, Lange, Ophoff \& Sabatti}]{ZhangLange:2010}
\textsc{Zhang, Z.}, \textsc{Lange, K.}, \textsc{Ophoff, R.} \& \textsc{Sabatti,
  C.} (2010).
\newblock Reconstructing {DNA} copy number by penalized estimation and
  imputation.
\newblock \textit{Ann. Appl. Stat.} \textbf{4}, 1749--1773.

\end{thebibliography}
\end{document}